\documentclass[12pt, a4paper]{article}
\usepackage[absolute]{textpos}
\usepackage[top=3cm,bottom=3cm,left=2cm,right=2cm]{geometry}
\usepackage{amsmath,amssymb,mathtools,dsfont,emptypage,graphicx,dcolumn,bm,booktabs,colortbl,collcell,enumerate,float,verbatim,multirow,xparse,slashed,mathrsfs,color}
\usepackage{cite}
\usepackage{braket}
\usepackage{tabularx}
\usepackage{multirow}
\usepackage{graphicx}
\usepackage{subcaption}
\usepackage{appendix}
\usepackage[labelfont=bf, labelsep=period]{caption}
\usepackage{sectsty}
\sectionfont{\fontsize{15}{12}\selectfont}
\subsectionfont{\fontsize{13}{12}\selectfont}

\usepackage[compat=1.1.0]{tikz-feynman}
\usepackage{contour}

\definecolor{myred}{cmyk}{0,1,1,0.55}
\definecolor{mygreen}{rgb}{0.27, 0.64, 0.48}
\definecolor{mygray}{gray}{.95}
\usepackage[colorlinks=true,urlcolor=myred,linkcolor=black,citecolor=mygreen]{hyperref}

\begin{document}


\begin{center}
{\bf \large Standard and Non-Standard Aspects of Neutrino Physics} \\
[5mm]
\renewcommand*{\thefootnote}{\fnsymbol{footnote}}
Alessandro Granelli \footnote{\href{mailto:alessandro.granelli@unibo.it}{alessandro.granelli@unibo.it}}
\\
\vspace{2mm}
{\it Dipartimento di Fisica e Astronomia, Università di Bologna, via Irnerio 46, 40126 and} \\
{\it INFN, Sezione di Bologna, viale Berti Pichat 6/2, 40127, Bologna, Italy.}
\end{center}

\begin{center}
    {\bf Abstract}
\end{center}
This review provides a succinct overview of the basic aspects of neutrino physics. The topics covered include: neutrinos in the standard model and the three-neutrino mixing scheme; the current status of neutrino oscillation measurements and what remains to be determined; the seesaw mechanisms for neutrino mass generation and the associated phenomenology, including the leptogenesis mechanism to explain the observed matter-antimatter asymmetry of the Universe; models for the origin of the pattern of neutrino mixing and lepton masses based on discrete flavour symmetries and modular invariance.\\
\hrule
\vspace{.5em}

\renewcommand*{\thefootnote}{\arabic{footnote}}
\setcounter{footnote}{0}


\section{Introduction}
Since W.~Pauli's original proposal of the neutrino idea 
\cite{Pauli:83282}, E.~Fermi's first description of the weak force
\cite{Fermi1934}, and the pioneering antineutrino detection with the Cowan-Reines experiment \cite{Reines:1956rs, Cowan_1956}, our understanding of neutrino properties and their interactions has continuously expanded. Considerable progress has been made from the experimental side, enabling the detection of neutrinos from multiple sources: the Sun, Supernovae, the interior of the Earth, radioactive nuclei and reactor plants, cosmic rays interacting with the atmosphere, and, tentatively, from extragalactic sources such as active galactic nuclei (see, e.g., \cite{Vitagliano:2019yzm} for a complete description of neutrino fluxes across different energies). Such outstanding experimental advancements, along with the well-established electromagnetic and cosmic rays observation techniques and the advent of gravitational wave detectors \cite{LIGOScientific:2016sjg}, signal a promising era of multi-messenger astronomy.

In parallel, a series of theoretical developments 
 ultimately led to the formulation of the \textit{Standard Model} (SM) of particle physics \cite{Glashow:1961tr, Weinberg:1967tq, Salam:1968rm}. The SM
 represents, at present, the most accurate description of the known fundamental particles, including neutrinos, and forces, with the exception of gravity. 
 However, despite its great successes in predicting how Nature behaves at the sub-atomic scales, 
 the SM remains incomplete. One motivation is that the SM assumes neutrinos are massless. However, the observed phenomenon of neutrino oscillations \cite{SNO:2002tuh, Super-Kamiokande:1998kpq} can be explained by the standard paradigm of the \textit{three-neutrino mixing scheme}, which considers at least two massive neutrinos and the mixing among three neutrino species. This scheme cannot be incorporated into the SM theory without interfering with some of its fundamental properties. Besides, the SM falls short in describing other challenging issues, including the long-standing problems of explaining the nature of dark matter \cite{Arbey:2021gdg} and the overabundance of matter over antimatter in the Universe \cite{DiBari:2021fhs}. These fundamental problems may be inter-related, with neutrinos potentially playing a primary role in establishing such connections.
 
The search for new physics beyond the SM is a central focus in high-energy (astro)particle physics. Numerous extensions of the SM have been proposed to explain the origin of neutrino masses while attempting to tackle other fundamental problems in modern physics. The phenomenology of many of these extensions can be tested through the extensive experimental programs that are underway. Amidst the renewed excitement and recent advances in the field, this short review aims at gathering concisely some of the standard and non-standard aspects of neutrino physics. Several reviews dedicated to specific topics treated in this work already exist in the literature, to which we will refer throughout the text. This work is more oriented on the particle physics side and related phenomenology, although neutrinos are relevant also in cosmology and astrophysics (see, e.g., \cite{Giunti:2007ry, Lesgourgues:2013sjj} for monographs on the topic). 

The manuscript is organised as follows: in Sec.~\ref{sec:StdnuPic} we briefly overview the physics of neutrinos in the SM and in the three-neutrino mixing scheme; in Sec.~\ref{sec:CurrMeas} we summarise our current knowledge of the neutrino oscillation parameters and what is left to determine; in Sec.~\ref{sec:MassOrigin} we describe possible mechanisms for neutrino mass generation, concentrating on the seesaw models; in Sec.~\ref{sec:LepmixOrigin} we briefly discuss models for the origin of the lepton mixing pattern; we conclude in Sec.~\ref{sec:Summary}.

\section{The standard neutrino picture}\label{sec:StdnuPic}

\subsection{Neutrinos in the Standard Model of particle physics}
There are three kinds, or \textit{flavours}, of (anti)neutrinos, namely $\nu_e$, $\nu_\mu$ and $\nu_\tau$ ($\bar{\nu}_e$, $\bar{\nu}_\mu$ and $\bar{\nu}_\tau$). The flavour (anti)neutrino $\nu_\alpha$ ($\bar{\nu}_\alpha$), $\alpha = e,\,\mu,\,\tau$, is produced with a charged antilepton $\alpha^+$ (lepton $\alpha^-$), or produces a charged (anti)lepton $\alpha^-$ ($\alpha^+$) in Charged Current (CC) weak interaction processes. Hitherto, neutrinos (antineutrinos) have exhibited a polarisation in a direction that is opposite (parallel) to their motion, or, equivalently, with negative (positive) helicity. To account for such experimentally observed neutrino properties, neutrinos $\nu_\alpha$ and antineutrinos $\bar{\nu}_\alpha$ of lepton flavour $\alpha = e,\,\mu,\,\tau$ are described in the SM with a left-handed (LH) chiral field $\nu_{\alpha L}(x)$. In the absence of neutrino masses, this field destroys (creates) neutrinos (antineutrinos) with negative (positive) helicity. In the SM, neutrinos are considered massless, and no right-handed (RH) neutrino chiral field is included in its content as they would be singlet under the SM gauge group. As such, the corresponding RH neutrinos would be completely inert. Together with the LH charged lepton chiral fields $\alpha_L(x)$, the LH neutrino fields form SU(2) doublets $\psi_{\alpha L}(x) = (\nu_{\alpha _L}(x), \alpha_L(x))$ with hypercharge $+1$. 

The SM Lagrangian is symmetric under the action of global $\text{U}(1)_{L_\alpha}$, $\alpha = e,\,\mu,\,\tau$, and $\text{U}(1)_{L}$ transformation. In particular, the lepton flavour $L_\alpha$ assigned to (anti)leptons $\beta^-$ ($\beta^+$), $L_\alpha(\beta^{-}) = L_\alpha({\nu}_{\beta}) = \delta_{\alpha\beta}$ and $L_\alpha(\beta^{+}) = L_\alpha(\bar{\nu}_{\beta}) = -\delta_{\alpha\beta}$, and the total lepton number $L = L_e + L_\mu + L_\tau$ are conserved by the electroweak interactions.  However, these symmetries are accidental and arise solely because of the particle content and renormalisability of the theory.

The CC and Neutral Current (NC) electroweak interactions between leptons are described by the following terms in the SM Lagrangian:
\begin{equation}
\mathcal{L}^\text{SM}_\text{CC}(x) = -\frac{g_w}{2\sqrt{2}}j^\mu_W(x) W_\mu(x) + \text{h.~c.}\,,\quad
\mathcal{L}^\text{SM}_\text{NC}(x) = -\frac{g_w}{2c_\text{W}}j^\mu_Z(x)Z_\mu(x)\,,
\end{equation}
where $g_w\simeq 0.65$ 
is the $\text{SU}(2)_\text{L}$ weak coupling constant; $c_\text{W} \equiv \cos\theta_\text{W}\simeq 0.88$ 
($s_\text{W}\equiv \sin\theta_\text{W}\simeq 0.48$), 
and $\theta_\text{W}$ is the Weinberg angle; $W_\mu(x)$ and $Z_\mu(x)$ are the 4-components vector fields describing the $W^\pm$ and $Z^0$ weak massive bosons having masses $M_W \simeq 80.4\,\text{GeV}$ and $M_Z \simeq 91.2\,\text{GeV}$, respectively. 
The weak currents are defined as
\begin{eqnarray}\label{eq:NC_current}
j_W^\mu(x) &=& 2 \sum_{\alpha = e,\,\mu,\,\tau} \overline{\nu_{\alpha L}}(x)\gamma^\mu \alpha_L(x)\,,\\
        j_Z^\mu(x) &=& 2 \sum_{\alpha = e,\,\mu,\,\tau} \left[g_L^\nu\,\overline{\nu_{\alpha L}}(x)\gamma^\mu \nu_{\alpha L}(x) + g_L^\alpha \,\overline{\alpha_L}(x)\gamma^\mu\alpha_L(x) + g_R^\alpha\, \overline{\alpha_R}(x)\gamma^\mu\alpha_R(x)\right]\,,
        \label{eq:CC_current}
\end{eqnarray}
where $\gamma^\mu$, $\mu = 0,\,1,\,2,\,3$, are the four Dirac gamma matrices; $\alpha_{R}(x)$ is the RH chiral field for the charged lepton $\alpha$; the couplings take the values $g_L^\nu = +1/2$, $g_L^e = g_L^\mu = g_L^\tau = -1/2 + s_\text{W}^2$ and $g_R^e = g_R^\mu = g_R^\tau = s_\text{W}^2$. Quarks also participate to the CC and NC electroweak interactions, albeit with different NC coefficients. The parameters of the electroweak interaction are determined at present with great accuracy thanks to precision electroweak measurements and global fits. A complete list with the values of the relevant constants, couplings and masses can be found, e.g., in \cite{PDG_Workman:2022ynf}.

Neutrinos interact with ordinary matter via the CC and NC electroweak currents. These interactions regulate both the production and the detection of neutrinos in all the different sources and experimental facilities. Consequently, determining the scattering cross sections of the different neutrino scattering interactions at relevant energies is a crucial task in neutrino physics (see, e.g., \cite{Formaggio:2012cpf} for a comprehensive review on neutrino cross sections across multiple energy scales).

\subsection{The three-neutrino mixing scheme and neutrino oscillations} \label{sec:numixing}

It has been widely established by experiments investigating solar, atmospheric, reactor and accelerator neutrinos that flavour neutrinos undergo oscillations: an initial flavour neutrino $\nu_\alpha$ (antineutrino $\bar{\nu}_\alpha$), after travelling a certain distance for a period of time $t$, can
be detected as a flavour neutrino $\nu_\beta$ (antineutrino $\bar{\nu}_\beta$) of different flavour $\beta\neq \alpha$ with a non-zero probability $P(\nu_\alpha \to \nu_{\beta}, t)\neq 0$ ($P(\bar{\nu}_\alpha \to \bar{\nu}_{\beta}, t)\neq 0$). The three-neutrino mixing scheme explains the origin of such phenomenon by assuming that (at least two) neutrinos are massive and are a mixture of the three lepton flavour states, namely \cite{PDG_Workman:2022ynf}:
$
 \nu_{\alpha L}(x)=\sum_{a=1}^3 U_{\alpha a} \nu_{a L}(x),
$ 
%
where $\nu_{a L}(x)$, $a=1,2,3$, is the LH component of the field of a light neutrino $\nu_a$ with mass $m_a$,  
and $U$ is the $3\times 3$ unitary Pontecorvo-Maki-Nakagawa-Sakata (PMNS) 
neutrino (lepton) mixing matrix \cite{Pontecorvo:1957cp, Pontecorvo:1957qd, Maki:1962mu}. 
Adopting the standard parametrisation and numbering of the massive neutrinos, the PMNS matrix can be written as \cite{ParticleDataGroup:2014cgo}
\begin{equation}
\label{eq:PMNS}
U = \begin{pmatrix}
c_{12}c_{13}&s_{12}c_{13}&s_{13}e^{-i\delta}\\
-s_{12}c_{23}-c_{12}s_{23}s_{13}e^{i\delta}&c_{12}c_{23}-s_{12}s_{23}s_{13}e^{i\delta}&s_{23}c_{13}\\
s_{12}s_{23}-c_{12}c_{23}s_{13}e^{i\delta}&-c_{12}s_{23}-s_{12}c_{23}s_{13}e^{i\delta}&c_{23}c_{13}
\end{pmatrix}\times
\begin{pmatrix}
1&0&0\\
0&e^{\frac{i\alpha_{21}}{2}}&0\\
0&0&e^{\frac{i\alpha_{31}}{2}}
\end{pmatrix},
\end{equation}
%
with $c_{ab} \equiv \cos\theta_{ab}$, $s_{ab} \equiv \sin\theta_{ab}$ and
the angles $\theta_{ab} \in [0,\pi/2]$, with $a,\,b=1,\,2,\,3$;
$\delta\in[0,2\pi]$ is the Dirac phase;
$\alpha_{21}$ and $\alpha_{31}$ are the two Majorana phases \cite{Bilenky:1980cx}, 
$\alpha_{21 (31)} \in [0,2\pi]$. 

The phenomenon of flavour neutrino oscillations is a fascinating manifestation of quantum mechanics on macroscopic scales. Derivations of the probabilities $P(\nu_\alpha\to\nu_\beta, t)$ can be found in textbooks (see, e.g., \cite{Giunti:2007ry}). Here, we provide a concise description. A flavour state $|\nu_\alpha\rangle$ is a coherent superposition of massive states, such that, at time $t=0$, $|\nu_\alpha(t=0)\rangle = \sum_a  U^*_{\alpha a}|\nu_a(t=0)\rangle$. Under the plane-wave approximation, the initial massive states have definite spatial momentum $p_\nu\equiv |\vec{p}_\nu|$ and energy $E_\nu\simeq p_\nu$, with $E_a-E_{b} \simeq (m_a^2-m_{b}^2)/2E_\nu$, $a,\,b = 1,\,2,\,3$. The initial massive state is evolved unitarily in time through the Schr\"{o}dinger equation and later sandwiched with a flavour neutrino state $|\nu_\beta\rangle$. The final result of the calculation in the vacuum is
 \begin{equation}
    P(\nu_\alpha\to\nu_\beta, t) = |\langle \nu_\beta|\nu_\alpha(t)\rangle|^2 
    \simeq\left|\,\sum_{a = 1}^3 U^*_{\alpha a}U_{\beta a} \,\text{exp}\left(-i\frac{\Delta m_{a1}^2 L}{2E_\nu}\right)\right|^2,
\end{equation}
where $L\simeq t$ is the distance travelled by the initial relativistic neutrino, while $\Delta m_{ab}^2 \equiv m_{a}^2-m_{b}^2$, $a,\,b = 1,\,2,\,3$.
The expression for antineutrinos can be obtained by taking the conjugate of the PMNS matrix entries. Approaches with different levels of refinement, considering a superposition of wave-packets or in the quantum field theory formalism, yield the same results in all experimentally relevant situations \cite{Kayser:1981ye, Kiers:1995zj, Akhmedov:2009rb, Akhmedov:2010ms}. More importantly, interactions with the constituents of the medium in which neutrinos propagate affect the neutrino oscillation probability through \textit{matter effects} \cite{Wolfenstein:1977ue}. The Hamiltonian of the system is modified by the presence of an effective matter potential associated to the CC and NC weak interactions. Matter effects differ between constant and varying density conditions. The constant density case can be studied analytically and is relevant for accelerator experiments with long-baseline facilities \cite{Freund:1999gy, Freund:2001pn, Akhmedov:2004ny}. The varying density case, instead, is more subtle, but crucial in modelling, e.g., the propagation of solar neutrinos with energies of a few MeV. An adiabatic change of density in the Sun's core induces the Mikheyev–Smirnov–Wolfenstein (MSW) effect \cite{Wolfenstein:1977ue, Mikheyev:1985zog}, leading to a $\nu_e$ survival probability of $P(\nu_e\to\nu_e, t)\simeq \sin^2\theta_{12}$. 

\section{Current measurements and unknowns in the standard neutrino mixing paradigm}\label{sec:CurrMeas}
The parameters that enter the expressions of the neutrino oscillations probabilities are: the three mixing angles $\theta_{12}$, $\theta_{23}$ and $\theta_{13}$; the Dirac phase $\delta$; two independent squared mass difference denoted as $\Delta m_\odot^2\equiv \Delta m_{21}^2$ and $\Delta m_\text{atm}^2$, which, together with $\theta_{12}$ and $\theta_{23}$, drive the oscillations of solar and atmospheric neutrinos, respectively. The parameters $\theta_{12}$, $\theta_{23}$, $\theta_{13}$, $\Delta m_{21}^2$ and $|\Delta m_\text{atm}^2|$ have been determined with remarkably high precision by solar neutrino experiments \cite{Cleveland:1998nv, SNO:2011hxd, Kaether:2010ag, SAGE:2009eeu,  Super-Kamiokande:2023jbt}, 
atmospheric neutrino experiments \cite{IceCubeCollaboration:2023wtb, Super-Kamiokande:2023ahc}, 
reactor neutrino experiments with either short-baseline (SBL) \cite{DoubleChooz:2014kuw, DayaBay:2018yms, RENO:2018dro} 
and long-baseline (LBL) \cite{KamLAND:2013rgu}  
facilities, and in LBL accelerator-based neutrino experiments \cite{K2K:2006yov, MINOS:2013utc, OPERA:2018nar, T2K:2017hed, T2K:2018rhz, NOvA:2017ohq, NOvA:2018gge}.  Measurements of $\delta$ have also been reported by LBL accelerator experiments, although the uncertainty in its determination is still relatively large (see Sec.~\ref{sec:CPV}).

Depending on the experimental setup, each neutrino oscillation experiment is most sensitive to specific sets of neutrino mixing parameters. 
Global fit analyses conducted by various collaborations \cite{deSalas:2020pgw, Esteban:2020cvm, Capozzi:2021fjo} combine data from the different experiments on neutrino oscillations and provide an overall picture of the neutrino mixing pattern. For instance, the most recent \texttt{NuFit 5.2} global analysis \cite{nufit,
Esteban:2020cvm}
reported the following best-fit $\pm 1\sigma$ values and $3\sigma$ ranges of the mixing parameters in the case of a neutrino mass spectrum with normal (inverted) ordering (see further in Sec.~\ref{sec:ordering}): 
\begin{eqnarray}
    \nonumber \theta_{12}&=&33.41^\circ{}^{+0.75^\circ}_{-0.72^\circ},[31.31^\circ, 35.74^\circ];\\
    \nonumber
    \theta_{13} &=& 8.58\pm0.11^\circ\,(8.57^\circ{}\pm0.11^\circ), [39.7^\circ, 51.0^\circ] ([39.9^\circ, 51.5^\circ]);\\
    \nonumber
    \theta_{23} &=& 42.2^\circ{}^{+1.1^\circ}_{-0.9^\circ}(49.0^\circ{}^{+1.0^\circ}_{-1.2^\circ}), [8.23^\circ, 8.91^\circ] ([8.23^\circ, 8.94^\circ]);\\
    \nonumber
  \Delta m_\odot^2/(10^{-5}\,\text{eV}^{2}) &=& 7.41^{+0.21}_{-0.20}, [6,82, 8.03];\\
  \nonumber
  \Delta m_\text{atm}^2/(10^{-3}\,\text{eV}^{2}) &=& 2.507^{+0.026}_{-0.027}\,(-2.486^{+0.025}_{-0.028}), [2.427, 2.590] ([-2.570, -2.406]).
 \end{eqnarray}

Despite the significant progress we have made in determining the neutrino mixing parameters, much remains to be understood. While future experiments will undoubtedly seek for greater precision, efforts will be dedicated to addressing the open problems that we outline in the subsequent subsections. 

\subsection{The ordering of the neutrino mass spectrum and the absolute mass scale}\label{sec:ordering}

All the existing neutrino data collected so far do not permit to infer conclusively the sign of $\Delta m^2_\text{atm}$. Correspondingly, there are two possible orderings for the neutrino mass 
spectrum that are consistent with current neutrino mixing data: 
\begin{itemize}
\item {\bf Normal Ordering (NO):} $m_1< m_2< m_3$,
$\Delta m_{31}^2 \equiv \Delta m_\text{atm}^2>0$;
\item {\bf Inverted Ordering (IO):} $m_3< m_1< m_2$,
$\Delta m_{32}^2\equiv \Delta m_\text{atm}^2<0$.
\end{itemize}
 Depending on the value of the lightest neutrino mass,
the neutrino mass spectrum can also be hierarchical, that is $0\simeq m_1 \ll m_2 < m_3$, with 
$m_2 \simeq (\Delta m_{21}^2)^{1/2}$ and 
$m_3 \simeq (\Delta m_{31}^2)^{1/2}$ for NO, or $0\simeq m_3 \ll m_1 < m_2$, 
with $m_1 \simeq (|\Delta m_{32}^2| - \Delta m_{21}^2)^{1/2}$ and 
$m_2 \simeq |\Delta m_{32}^2|^{1/2}$ for IO. The possibility of a quasi-degenerate spectrum with $m_1\simeq m_2\simeq m_3$, 
with $m^2_{1,2.3}\gg |\Delta m^2_{31(32)}|$ is also not excluded by current data. The global fit analyses on oscillation data only \cite{deSalas:2020pgw, Esteban:2020cvm, Capozzi:2021fjo}, as well as with the inclusion of data on $\beta$-decay and from cosmological probes \cite{Gariazzo:2022ahe}, indicate a preference of NO over IO up to only $\sim 2.7\sigma$ confidence level (C.~L.). 

Since $\theta_{13}$ is relatively large, more information on the sign of $\Delta m_\text{atm}^2$ can be obtained in the future from LBL accelerator facilities (NO$\nu$A \cite{NOvA:2004blv, NOvA:2017ohq, NOvA:2018gge}, DUNE \cite{DUNE:2021tad}) and atmospheric neutrino experiments (INO \cite{ICAL:2015stm}, PINGU \cite{IceCube-PINGU:2014okk, IceCube:2016xxt}, ORCA \cite{KM3NeT:2021ozk}, DUNE \cite{DUNE:2021tad}
and Hyper-Kamiokande (HK) \cite{Hyper-Kamiokande:2018ofw}) exploiting matter effects. Furthermore, based on a different method \cite{Petcov:2001sy}, the reactor experiment JUNO \cite{JUNO:2015zny, JUNO:2021vlw} is also designed to determine with precision the neutrino mass ordering. If neutrinos are of the Majorana type, information on the mass ordering can also be attained from experiments on neutrinoless double beta ($(\beta\beta)_{0\nu}$-)decay (see Sec.~\ref{sec:DorM} for more details). 

The absolute mass scale of neutrinos, i.e., the value of $m_{1(3)}$ in the NO (IO) case, is currently not known and only upper bounds are available. The ongoing KATRIN experiment \cite{Wolf:2008hf} is aimed at measuring the neutrino mass by examining the tail of the electron spectrum in the final state of $\beta$-decay of tritium ${}^3_1\text{H}$ and provides the following limit \cite{Aker:2019uuj,Aker:2021gma}:
$m_{1,2,3} < 0.8\,\text{eV}$ ($90\%$ C.L.). From the best existing lower bounds on the $(\beta\beta)_{0\nu}$-decay lifetimes of $^{136}$Xe \cite{KamLAND-Zen:2016pfg} and $^{76}$Ge \cite{Agostini:2020xta} one can obtain a bound on neutrino masses, which, however, applies only if neutrinos have a Majorana nature \cite{Penedo:2018kpc}:
$m_{1,2,3} \lesssim 0.58$ eV. The most stringent upper bound on the sum of neutrino masses comes from observations of the Cosmic Microwave Background by PLANCK \cite{Planck:2018vyg}, combined with
supernovae and other cosmological and astrophysical data, yielding $\sum_a m_a < (0.12-0.69)\,\text{eV}$ (95$\%$ C.L.) depending on the adopted cosmological model and level of statistical complexity~\cite{LVinZyla:2020zbs, Capozzi_2020} (see also \cite{Vagnozzi2017, RoyChoudhury:2019hls, Ivanov:2019hqk, DES:2021wwk, Tanseri2022}).

\subsection{CP-violation in the PMNS matrix}\label{sec:CPV}
Whether there is CP-violation in the lepton sector originating from the Dirac phase and/or the Majorana phases in the PMNS matrix is still not confidently known.
The requisite for the CC interaction term of being symmetric under CP transformation restricts the PMNS matrix entries to be either purely real or imaginary, $U_{\alpha a} = \pm U_{\alpha a}^*$, a condition which is satisfied when the Dirac and Majorana phases take the CP-conserving values $\delta = k\pi$, $\alpha_{21} = k_{21}\pi$, $\alpha_{31} = k_{31}\pi$, with $k, \,k_{21},\,k_{31} = 0,\,1,\,2$ (see, e.g., \cite{Pascoli:2006ci}). If $\delta \neq k\pi$ and/or, provided that neutrinos are of the Majorana type (see Sec.~\ref{sec:DorM}), $\alpha_{21} \neq k_{21}\pi$ and/or $\alpha_{31} \neq k_{31}\pi$, there is CP-violation in the lepton sector from the phases of the PMNS matrix.

The Dirac phase $\delta$ is responsible for CP-violating effects in neutrino oscillations. Such effects can be computed as the differences between the oscillation probabilities $P(\nu_\alpha \to \nu_{\beta}, t)- P(\bar{\nu}_\alpha \to \bar{\nu}_{\beta}, t)$, which results to be proportional to the invariant $J_\text{CP} = s_{12}c_{12}s_{23}c_{23}s_{13}c_{13}^2\sin \delta$ \cite{Krastev:1988yu}.
There have been some indications of CP-violation in neutrino oscillations from the LBL experiments Tokai-to-Kamioka (T2K) and NO$\nu$A, although with some tensions (a review of their results and discrepancies is given in, e.g., \cite{Rahaman:2022rfp}). However, global fit analyses cannot confidently exclude CP-conserving values yet. E.g., the \texttt{Nufit 5.2} analyses determined $
  \delta = 232^\circ {}^{+36^\circ}_{-26^\circ} (276^\circ{}^{+22^\circ}_{-29^\circ}), [144^\circ, 350^\circ] ([194^\circ, 344^\circ]),
$ with a relatively large uncertainty. Future determination are planned for the upcoming LBL accelerator experiments T2K \cite{T2K:2011ypd, T2K:2017hed, T2K:2018rhz} , NO$\nu$A \cite{NOvA:2004blv, NOvA:2017ohq, NOvA:2018gge}, Tokai-to-HK (T2HK) \cite{Hyper-Kamiokande:2018ofw} and DUNE \cite{DUNE:2021tad} (see also the Tokai-to-HK-to-Korea (T2HKK) proposal \cite{Hyper-Kamiokande:2016srs, Panda:2022vdw}).

In contrast, the Majorana phases do not enter the flavour neutrino oscillation probabilities \cite{Bilenky:1980cx,Langacker:1986jv}, 
but contribute to the $(\beta\beta)_{0\nu}$-decay rate 
(see Sec.~\ref{sec:DorM}).

\subsection{Neutrinos can be either Dirac or Majorana fermions}\label{sec:DorM}
A massive spin-$1/2$ particle $\psi$ with mass $m_\psi$ can either be of the Dirac or Majorana type if it is, respectively, distinguishable or indistinguishable from its corresponding antiparticle $\overline{\psi}$. Since all the known fundamental fermions are electrically charged, except neutrinos, they are intrinsically of the Dirac type. In quantum field theory terms, a Dirac particle $\psi$ can be described by a 4-component spinor $\psi_D(x)$ that satisfies the Dirac equation.
After quantisation, the field $\psi_D(x)$ annihilates (creates) particles $\psi$ (antiparticles $\overline{\psi}$) having helicities with either sign, thus describing in total four degrees of freedom. The field $\psi_D(x)$ can be decomposed in terms of LH and RH independent chiral fields $\psi_L(x)$ and $\psi_R(x)$, such that $\psi_D(x) = \psi_L(x) + \psi_R(x)$. The mass term for a Dirac field would be:
\begin{equation}
    -\mathcal{L}_\text{Dirac}^m(x) = \overline{\psi_L}(x)m_\psi\psi_R(x) + \text{h.~c.} = \overline{\psi_D}(x)m_\psi \psi_D(x).
    \end{equation}
Any $\text{U}(1)$ symmetry acting on the Dirac field as $\psi_D(x)\to e^{iq}\psi_D(x)$, $q>0$, leaves the Dirac mass term invariant. 

A Majorana particle $\psi = \overline{\psi}$ can be described with a 4-component spinor $\psi_M(x)$ which both satisfies the Dirac equation and the Majorana condition
$
\psi_M^c(x) = C(\overline{\psi_M}(x))^T
$, 
where $C$ is the charge conjugation matrix \footnote{The Majorana condition can be defined up to an overall unphysical phase which we have taken equal to unity.}. The Majorana condition is invariant under a $\text{U}(1)$ transformation acting on the Majorana field as $\psi_M(x) \to e^{iq}\psi_M(x)$ only if $q=0$. In this sense, Majorana fields can only describe neutral particles.
The Majorana condition restricts the number of degrees of freedom to two, and the quantised Majorana field $\psi_M(x)$ creates and annihilates particles $\psi$ with either positive or negative helicity. As such, a Majorana field $\psi_M(x)$ can be decomposed in terms of its LH (or the RH) component $\psi_L(x)$ only, namely $\psi_M(x) = \psi_L(x) + \psi_R^c(x)$, with $\psi_R^c = C(\overline{\psi_L}(x))^T$. The mass term of a Majorana field takes the form
\begin{equation}
     -\mathcal{L}_\text{Majorana}^m(x) =\frac{1}{2}\overline{\psi^c_R}(x)m_\psi \psi_{L}(x)+ \text{h.~c.} = \frac{1}{2}\overline{\psi_M}(x)m_\psi\psi_M(x).
 \end{equation}
and breaks any $\text{U}(1)$ transformation acting on $\psi_L(x)$ by two charge units. \footnote{More details on the basic properties of Dirac and Majorana fermions in quantum field theory can be found, e.g., in \cite{Pal:2010ih} (see also \cite{Bilenky:1987ty}).}
 
The evidence of non-vanishing neutrino masses gives rise to the question whether neutrinos are of the Dirac or Majorana type. 
The nature of neutrinos, either Dirac or Majorana, is intimately related to the conservation/breaking of total lepton number. Dirac neutrinos would conserve $\text{U}(1)_L$ and the associated total lepton charge, while Majorana neutrinos would break it by two units. One of the consequences of this is that the Majorana phases $\alpha_{21}$ and $\alpha_{31}$ in the PMNS matrix of Eq.~\eqref{eq:PMNS} (can) cannot be reabsorbed into a redefinition of the charged lepton fields and the Majorana (Dirac) neutrino fields. This means that the Majorana phases $\alpha_{21}$ and $\alpha_{31}$ can only play a role in physical processes in which the Majorana nature of neutrinos is manifest, while they do not contribute to any physical observable if neutrinos are of the Dirac type. In both cases, the flavour lepton number is inevitably broken by neutrino oscillations.

Whether neutrinos are of the Dirac or Majorana type can be established experimentally by searching for lepton number violating processes. The most promising searches are those looking for the $(\beta\beta)_{0\nu}$-decay, that is the double $\beta$-decay of a nucleus without the emission of neutrinos. This process would be a direct manifestation of the Majorana nature of neutrinos (see, e.g., \cite{Agostini:2022zub} for a recent review). The decay rate of the $(\beta\beta)_{0\nu}$-decay, in the minimal case, is proportional to a nuclear matrix element $|\mathcal{M}_N|^2$ which contains most of the uncertainty, and the effective Majorana mass parameter
\begin{equation}
|m_{\beta\beta}|^2 = \left|m_1 |U_{e1}|^2 + m_2 |U_{e2}|^2 e^{i\alpha_{21}}+ m_3 |U_{e3}|^2 e^{i(\alpha_{31}-2\delta)}\right|^2,
\end{equation}
which depends on the neutrino masses, the PMNS mixing angles and the Dirac and Majorana phases. Through the mass dependence, the predictions of the effective Majorana mass parameter vary depending on the neutrino mass spectrum. The experiments that are currently searching for the $(\beta\beta)_{0\nu}$-decay, as well as proposals for the future, are several (see, e.g., an extended list in \cite{Agostini:2022zub}), but, at present, no evidence has been found. The null detection allows putting constraints on the effective mass and the absolute mass scale, depending on the ordering and precision with which the nuclear matrix element is determined. The effective Majorana mass parameter also depend on the Majorana phases, even though it remains challenging to obtain valuable information in the near future \cite{Agostini:2022zub}.

\subsection{Deviations from the standard three-neutrino mixing scheme}
A pertinent aspect to understand is whether the three-neutrino mixing scheme concludes the narrative on neutrino oscillations or deviations from the standard picture should be considered \cite{Arguelles:2022tki}. Below, we briefly discuss a few examples of such deviations.

\subsubsection{Sterile neutrinos}
There could be (at least) one additional neutrino, $\nu_s$, that mixes with ordinary neutrinos. If the corresponding massive state, $\nu_4$, is sufficiently light, neutrinos of this kind can be produced through coherent oscillations with ordinary neutrinos. However, the number of neutrino species, $N_\nu$, participating in the electroweak interaction is constrained by precision measurements of the rate of the invisible decay $Z^0\to\nu_{\alpha}\bar{\nu}_\alpha$, yielding $N_\nu = 2.9963\pm 0.0074$ \cite{PDG_Workman:2022ynf, Voutsinas:2019hwu, Janot:2019oyi}. Furthermore, any additional species oscillating with the ordinary neutrinos should exhibit a suppressed mixing to avoid compromising current neutrino oscillation data. In practice, extra neutrinos $\nu_s$ should be almost entirely \textit{sterile}. 

A $(3+1)$-neutrino mixing scheme including a sterile neutrino $\nu_4$ with mass $m_4\gtrsim \,\text{eV}$ and a squared mass difference $\Delta m_\text{SBL}^2 \equiv m_4^2-m_1^2\gtrsim 1\,\text{eV}$ has been proposed as an explanation for the Liquid Scintillator Neutrino Detector (LSND) anomaly, the MiniBooNE low-energy excess, the Gallium Anomaly (GA) and the Reactor Antineutrino Anomaly (RAA), despite with some tensions. In light of recent analyses, the RAA seems to be diminished and the GA is being investigated with more care (see, e.g., \cite{Acero:2022wqg, Elliott:2023cvh, Zhang:2023zif} for recent reviews), while the SBL Neutrino programme \cite{MicroBooNE:2015bmn, Cianci:2017okw} is aimed at shedding more light on the LSND and MiniBooNE anomalies. Nevertheless, the issue is still unresolved and it would be remarkable if the idea of a $\sim$ eV sterile neutrino is proven to be correct in the future. In turn, a sterile neutrino at the keV scale would be a good dark matter candidate, and active searches are ongoing \cite{Boyarsky:2018tvu}. 

\subsubsection{Non-unitary neutrino mixing matrix}
One major effect of the mixing with additional neutrino species is that the PMNS matrix results to be non-unitary. Essentially, the full $N_\nu\times N_\nu$ matrix that regulates the mixing between all the neutrino species is unitary, but, when restricted to the $3\times 3$ block corresponding to the PMNS matrix, unitarity is lost. A non-unitary PMNS matrix can affect CC and NC interactions, neutrino oscillations, electroweak precision data and flavour observables, as well as the history of the Universe, resulting in bounds on such deviations \cite{Forero:2021azc, Gariazzo:2022evs,  Blennow:2023mqx}.

\subsubsection{Non-standard neutrino interactions}
Neutrinos could also interact with matter through non-standard interactions (NSI) \cite{Wolfenstein:1977ue, Guzzo:1991hi}, which can be of the CC or NC type and parameterised by the following 6-dimensional effective operators \cite{Farzan:2017xzy}:
\begin{eqnarray}
    \mathcal{L}_\text{CC}^\text{NSI}(x) &=& -2\sqrt{2}G_F \epsilon_{\alpha\beta}^{ff'L/R}(\overline{\nu_{\alpha L}}(x)\gamma_\mu \beta_L(x))(\overline{f'_{L/R}}(x)\gamma^\mu f_{L/R}(x)),\\
    \mathcal{L}_\text{NC}^\text{NSI}(x)&=& -2\sqrt{2}G_F \epsilon_{\alpha\beta}^{fL/R}\,~~(\overline{\nu_{\alpha L}}(x)\gamma_\mu \nu_{\beta L}(x))(\overline{f_{L/R}}(x)\gamma^\mu f_{L/R}(x)),
\end{eqnarray}
$\alpha,\,\beta = e,\,\mu,\,\tau$, where $G_F = g_w^2/(4\sqrt{2}M^2_W)$ is the Fermi coupling constant; $\epsilon_{\alpha\beta}^{ff'L/R}$ and $\epsilon_{\alpha\beta}^{fL/R}$ are the dimensionless quantities that set the strength of the NSI; $f_{L/R}(x)\neq f'_{L/R}(x)$ are the LH/RH chiral fields for quarks (or leptons) for the CC NSI (NC NSI). Neutrino NSI can arise in a variety of SM extensions \cite{Proceedings:2019qno}, with the form given being generated after integrating out some heavy mediator fields with a mass that is much heavier than the momentum transferred in neutrino interaction processes. The presence of neutrino NSI can interfere with the picture of neutrino oscillations in various ways, by affecting  detection/production and through variations of the matter effects when propagating in a medium. NSI can also induce degeneracy in the determination of the neutrino mixing parameters \cite{Farzan:2017xzy}. Current oscillation data, electroweak precision data and data on coherent elastic neutrino-nucleus scattering impose constraints on their couplings \cite{Farzan:2017xzy, Esteban:2018ppq, Coloma:2023ixt}. 

\section{Possible origins of neutrino masses beyond the Standard Model}\label{sec:MassOrigin}

\subsection{Dirac mass term from the usual Higgs mechanism}\label{sec:Diracmass}
A minimal extension of the SM that explains the origin of neutrino masses consists of adding $n_R = 2,\,3$ RH neutrinos (fields) $\nu_{jR}$ ($\nu_{R j}(x)$), $j=1,\,...,n_R$ that couple via Yukawa interaction with the charged leptons and Higgs doublets. Such coupling is allowed by the symmetry of the SM and reads:
\begin{equation}
    -\mathcal{L}_Y^\nu(x) = \lambda^\nu_{\alpha j} \overline{\psi_{\alpha L}}(x) i\sigma_2\Phi^*(x) \nu_{jR}(x) + \text{h.~c.}
\end{equation}
where $\sigma_2$ is the second Pauli matrix, $\Phi(x) = (\Phi^+(x), \Phi^{(0)}(x))$ is the Higgs doublet field. After the electroweak symmetry is spontaneously broken and the neutral component of the Higgs doublet acquires a non-vanishing vacuum expectation value (VEV) $v/\sqrt{2} \simeq 174\,\text{GeV}$, the Yukawa term above generates Dirac mass terms for the Dirac fields $\nu_a(x) = (U_L)_{\alpha a}^* \nu_{\alpha L}(x) + (U_R)^*_{aj}\nu_{jR}$, $a=1,\,...,\,n_R$, with $U_L$ and $U_R$ being, respectively $3\times 3$ and $n_R\times n_R$ unitary matrices that factorise $\lambda^\nu$ according to the singular value decomposition. The field $\nu_a(x)$ describes the massive Dirac neutrino $\nu_a$ with mass $m_a = (v/\sqrt{2})\lambda^\nu_{\alpha a}(U^*_L)_{\alpha a}(U_R)_{aj}$, $a = 1,\,...,n_R$.  In the basis in which the charged lepton Yukawa couplings are diagonal, $U_L$ coincides with the PMNS matrix, $U_L = U$, and regulates the mixing of the LH components of the Dirac neutrinos $\nu_a$ as in the three-neutrino mixing scheme. \footnote{We note that, in case $n_R=2$, one neutrino remains massless. This latter situation is viable as neutrino oscillations dictates the existence of at least two non-zero neutrino masses.}

 There are several reasons why this mechanism is not particularly appealing. Firstly, from a phenomenological perspective, it leads to a dead-end. The RH neutrinos are singlets under the SM gauge group and do not mix with the LH ones, with which they constitute Dirac pairs. As such, they do not participate in the SM interactions and are completely sterile. The only observable phenomenon beyond the SM in this extensions is that of ordinary neutrino oscillations. 
 From the theoretical standpoint, it seems artificial to justify tiny Yukawa couplings $\lambda^\nu\sim 10^{-12}$ to explain the smallness of the neutrino masses. Additionally, imposing a global $U(1)_L$ symmetry by hand to avoid a Majorana mass term for the RH neutrinos seems contrived. All these factors motivate the exploration of alternative mechanisms for neutrino mass generation.

\subsection{Majorana mass term from the Weinberg operator and seesaw mechanisms}
The SM can be augmented with the effective Weinberg operator \cite{Weinberg:1979sa}
\begin{equation}
    -\mathcal{L}_\text{Weinberg}(x) = \frac{c_{\alpha\beta}}{\Lambda}\left(\overline{\psi^c_{\alpha R}}(x)\sigma_2\Phi(x)\right)\left(\Phi^T(x)\sigma_2\psi_{\beta L}(x)\right)+\text{h.~c.}
\end{equation}
where $\psi_{\alpha R}^c(x) = C(\overline{\psi_{\alpha L}}(x))^T,$ $c_{\alpha\beta}=c_{\beta\alpha}$ are symmetric coefficients and $\Lambda$ has the dimension of a mass. The Weinberg operator is the only 5-dimensional operator that can be constructed out of the field content of the SM theory \cite{Grzadkowski:2010es}. With such term added, the SM loses its renormalisabilty. This means, in particular, that such an extension will not be valid at any scale. The limit of validity is given by $\Lambda$, setting the scale at which new physics beyond the SM manifests itself. After the breaking of the electroweak symmetry, the Weinberg operator leads to a Majorana mass terms for the Majorana fields $\nu_a(x) = U_{\alpha a}^*\nu_{\alpha L}(x) + U_{\alpha a}^*\nu_{\alpha R}^c(x)$, $\nu_a^c(x) = \nu_a(x)$, where $\nu_{\alpha R}^c = C(\overline{\nu_{\alpha L}}(x))^T$, $a=1,\,2,\,3$. The field $\nu_a(x)$ describes a Majorana neutrino $\nu_a$ with mass $m_a =v^2 c_{\alpha\beta}U^*_{\alpha a}U^*_{\beta a}/(2\Lambda)$, $a=1,\,2,\,3$, suppressed by the scale $\Lambda\gg v$.

\begin{figure}
\centering
\begin{tikzpicture}[baseline=(current bounding box.center)]
\begin{feynman}
\vertex(V1);
\vertex(V2)[right=1cm of V1];
\vertex(V3)[right=1.8cm of V2];
\vertex(V4)[right=1cm of V3];
\vertex(H1)[above=1.4cm of V2];
\vertex(H2)[above=1.4cm of V3];
\node(X)[right=1.78cm of V1, label={below:\(M_N\)}][crossed dot];
\diagram*{
(V2)--[fermion, edge label=\(\nu_{L}\)](V1),
(V2)--[plain, edge label=\(N_R\)](X)--[plain, edge label=\(N_R\)](V3),
(V3)--[fermion, edge label' = \(\nu_{L}\)](V4),
(V2) --[scalar, edge label=\(\langle\Phi\rangle\)](H1),
(V3) --[scalar, edge label'=\(\langle\Phi\rangle\)](H2)
};
\end{feynman}
\end{tikzpicture}\qquad\qquad
\begin{tikzpicture}[baseline=(current bounding box.center)]
\begin{feynman}
\vertex(V1);
\vertex(V2)[right=1.6cm of V1];
\vertex(V3)[right=1.6cm of V2];
\vertex(D)[above=.8cm of V2];
\vertex(H1)[above left=1.1cm of D];
\vertex(H2)[above right=1.1cm of D];
\diagram*{
(V2)--[fermion, edge label=\(\nu_{L}\)](V1),
(V2)--[fermion, edge label'=\(\nu_{L}\)](V3),
(V2)--[scalar, edge label' = \(\Delta^{(0)}\)](D),
(D) --[scalar, edge label=\(\langle\Phi\rangle\)](H1),
(D) --[scalar, edge label'=\(\langle\Phi\rangle\)](H2)
};
\end{feynman}
\end{tikzpicture}\qquad\qquad
\begin{tikzpicture}[baseline=(current bounding box.center)]
\begin{feynman}
\vertex(V1);
\vertex(V2)[right=1cm of V1];
\vertex(V3)[right=1.8cm of V2];
\vertex(V4)[right=1cm of V3];
\vertex(H1)[above=1.4cm of V2];
\vertex(H2)[above=1.4cm of V3];
\node(X)[right=1.78cm of V1, label={below:\(M_\Sigma\)}][crossed dot];
\diagram*{
(V2)--[fermion, edge label=\(\nu_{L}\)](V1),
(V2)--[plain, edge label=\(\Sigma_R^{(0)}\)](X)--[plain, edge label=\(\Sigma_R^{(0)}\)](V3),
(V3)--[fermion, edge label' = \(\nu_{L}\)](V4),
(V2) --[scalar, edge label=\(\langle\Phi\rangle\)](H1),
(V3) --[scalar, edge label'=\(\langle\Phi\rangle\)](H2)
};
\end{feynman}
\end{tikzpicture}
\caption{Diagrammatic representations of neutrino mass contributions in type-I (left), type-II (center), and type-III (right) seesaw mechanisms. Here, $\langle\Phi\rangle$ stands for the VEV of the Higgs boson, $\nu_L$ is the LH component of a massive neutrino, $N_R$ is the RH component of a singlet fermion, $\Delta^{(0)}$ is the neutral component of the scalar triplet and $\Sigma_R^{(0)}$ is the RH projection of the neutral component of a fermion triplet. The crosses in the fermion propagators signify mass insertions.}\label{fig:Seesaw_diag}
\end{figure}
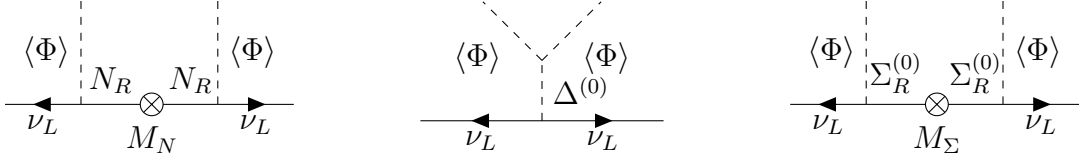

The Weinberg operator can be interpreted as the low-energy limit of a renormalisable interaction. This is in analogy with Fermi's theory being the low-energy limit of the CC weak interaction after integrating out the massive bosons $W^\pm$. There are only three possible 4-dimensional operators allowed by the SM symmetry that can generate the Weinberg operator at low-energy. These are classified on the basis of the type of mediator, which can be either one (or more) singlet fermion(s) $N$ with mass $M_N$ \cite{Minkowski:1977sc,Yanagida:1979as,GellMann:1980vs,Glashow:1979nm,Mohapatra:1979ia}, or $\text{SU}(2)$-triplet scalar(s) $\Delta$ with mass $M_\Delta$ \cite{Magg:1980ut, Schechter:1980gr, Mohapatra:1980yp}, or $\text{SU}(2)$-triplet fermion(s) $\Sigma$ with mass $M_\Sigma$ \cite{Foot:1988aq}. A specific mechanism of neutrino mass generation is associated to each of these operator, all having in common that the masses of the neutrinos get suppressed by the mass of the heavy mediator, metaphorically as sitting on a see-saw. For this reason, such mechanisms (models) for neutrino mass generation are referred to as \textit{type-I seesaw} for the singlet fermion, \textit{type-II seesaw} for the triplet scalar and \textit{type-III seesaw} for the triplet fermion (see, e.g., \cite{Cai:2017mow} for a review). The contributions to neutrino masses in the different seesaw mechanisms are schematised diagrammatically in Fig.~\ref{fig:Seesaw_diag}. In what follows, we give more details on each specific seesaw mechanism and related phenomenology \footnote{Majorana mass terms can also be generated radiatively at the one or more loop levels \cite{Zee:1980ai,Petcov:1982en,Petcov:1984nz, Babu:1988ig, Farzan:2012ev}.}.

\subsubsection{Type-I seesaw}
In the context of type-I seesaw mechanisms, the SM is augmented with $n_R$ RH neutrinos (fields) $N_{jR}$ ($N_{jR}(x)$), $j = 1,\,...,\,n_R$, taken to be singlet under the SM gauge group. 
In what follows, we concentrate on the case $n_R = 3$, even though the case with $n_R=2$ is the minimal scenario compatible with the existing neutrino oscillation data. The RH neutrinos enter in the SM Lagrangian with a Majorana mass term and a Yukawa coupling to the Higgs and LH charged lepton doublets:
\begin{equation}
\label{eq:Lseesaw}
-{\cal L}_\text{Type-I}(x) = \left(
\,Y_{\alpha j} \overline{\psi_{\alpha L}}(x)\,i\sigma_2\,\Phi^*(x)\,N_{jR}(x)+\hbox{h.c.}\right) + \frac{1}{2}M_j\overline{N_j}(x)N_j(x)\,,
\end{equation}
written in the basis for which both the charged lepton Yukawa and the Majorana mass matrix for the right-handed neutrinos are diagonal. The Majorana field $N_j(x) = N_{jR}(x) + N_{jL}^c(x)$, $N_{jL}^c(x) = C(\overline{N_{jR}}(x))^T$ describes a massive Majorana neutrino $N_j$ with mass $M_j>0$, $j=1,2,3$. \footnote{Typically, the values of $M_{1,2,3}$ are larger than the sub-eV scale of the light neutrino masses, and so we will refer to $N_{1,2,3}$ further on as \textit{heavy Majorana neutrinos}, or simply heavy neutrinos.} With the Higgs field acquiring a non-vanishing VEV, diagonalisation of the seesaw Lagrangian in Eq.~\eqref{eq:Lseesaw} leads to the type-I seesaw formula for the light neutrino masses:
$m_a \simeq -v^2U^*_{\alpha a}U^*_{\beta a}Y_{\alpha j}Y_{\beta j}/(2M_j)$,
which is valid at tree level and up to second-order corrections in the quantity $\Theta_{\alpha j} \equiv (v/\sqrt{2}) Y_{\alpha j}/M_j$, $|\Theta_{\alpha j}|\ll 1$. \footnote{See, e.g., \cite{Grimus:2000vj} for a derivation of the seesaw relations at arbitrary order in $|\Theta_{\alpha j}|$.} \footnote{One-loop corrections to light neutrino masses in the type-I seesaw scenario are discussed in, e.g., \cite{Grimus:2002nk, AristizabalSierra:2011mn, Lopez-Pavon:2012yda, Lopez-Pavon:2015cga}.}

The typical seesaw mechanism consists of setting $M_j$ at scales that are high enough to suppress the light neutrino masses. However, the smallness of the light neutrino masses can also be accounted for by tiny Yukawa couplings or cancellations occurring between the various terms (note that the seesaw formula is a sum of multiple terms). 
Consequently, the masses of the heavy neutrinos are basically unconstrained down to the $\sim \,\text{eV}$ scale when considered as an explanation for the light neutrino masses.
The latter situation can be justified when, e.g., two heavy neutrinos form a pseudo-Dirac pair \cite{Petcov:1982ya} and there is an approximately broken lepton number symmetry, as in \textit{inverse}, \textit{linear} and \textit{extended} seesaw scenarios (see, for instance, \cite{Abada:2015rta} for a related discussion).

At the same order in $|\Theta_{\alpha j}|$, the heavy neutrinos $N_j$ keep their masses $M_j$ and mix with the light neutrinos \footnote{Higher-order corrections lead, in particular, to deviations from unitarity of the PMNS matrix \cite{Blennow:2023mqx}.}:
\begin{equation}
\nu_{\alpha L}(x) \simeq U_{\alpha a} \nu_{aL}(x) + \Theta_{\alpha j} N_{jL}^c(x)\,.
\end{equation}
Through mixing with the light neutrinos, the heavy Majorana neutrinos inherit a coupling to the SM particles and participate in the CC and NC interactions. If their masses lie at the sub-TeV scale and their mixing is relatively large, they can be directly produced or detected at accelerator-based experiments and leave a variety of indirect low-energy signatures \cite{Atre:2009rg,Deppisch:2015qwa,Cai:2017mow}. An extensive experimental programme has been dedicated to the searches of heavy Majorana neutrinos.
The absence of evidence thus far imposes constraints on the mixing for masses ranging from the eV to the PeV scale, with additional bounds from cosmology and astrophysics \cite{Chrzaszcz:2019inj, Bolton:2019pcu, Urquia-Calderon:2022ufc}. Further experiments are planned and proposed for the future with the aim to probe the relevant parameter space more deeply (see, e.g., the reports \cite{Agrawal:2021dbo,Abdullahi:2022jlv, Antel:2023hkf}). Also future discussed colliders \cite{Antusch:2016ejd,FCC:2018evy,CEPCStudyGroup:2018ghi,
Li:2023tbx} have the potential to discover heavy Majorana neutrinos in the mass range from a few GeV to tens of TeV.

A common practice in phenomenological analyses is to express the Yukawa matrix in terms of low-energy observables via the Casas-Ibarra (CI) parameterisation \cite{Casas:2001sr}:
   \begin{equation}
   \label{eq:CasasIbarra}
   Y_{\alpha j} = 
   \pm i \frac{\sqrt{2}}{v} U_{\alpha a}\sqrt{m_a}O_{ja}\sqrt{M_j}\,,
   \end{equation}
 where the \textit{CI matrix} $O$ is an arbitrary $3\times 3$ complex orthogonal matrix that can be parameterised in terms of three complex angles (in the scenario of two heavy Majorana neutrinos, the number of complex angles reduces to one) and a phase that allows $\text{det}(O)=\pm 1$. In general, the CI matrix contains CP-violating phases in addition to those of the PMNS matrix.

\subsubsection{Type-II seesaw}
The minimal realisation of the type-II seesaw mechanism consists of extending the SM with a scalar $\text{SU}(2)_L$-triplet field $\Delta(x)$
given by
\begin{equation}
    \Delta(x) = \begin{pmatrix}
        \Delta^+(x)/\sqrt{2} & \Delta^{++}(x)\\
        \Delta^{(0)}(x) & -\Delta^+(x)/\sqrt{2}
    \end{pmatrix}.
\end{equation}
This field caries hypercharge $+1$ and enters the SM Lagrangian in the following lepton number violating terms:
\begin{equation}
-\mathcal{L}_\text{Type-II}(x) =  \frac{1}{2}Y^\Delta_{\alpha\beta}\overline{\psi_{\alpha R}^c}(x)\Delta(x)\psi_{\beta L}(x) + \mu_\Delta \Phi^T(x) i\sigma_2 \Delta^*(x) \Phi(x)+\text{h.~c.} 
\end{equation}
Here, $M_\Delta$ is the mass of the scalar triplet, $Y^\Delta_{\alpha\beta}$ are dimensionless couplings and $\mu_\Delta$ has the dimension of a mass. The scalar triplet $\Delta$ interacts with gauge bosons and, along with the Higgs doublet, contributes to the full scalar potential. We omit such terms for brevity. 

The masses of the light neutrinos are generated after both the Higgs and scalar triplets acquire non-vanishing VEVs, the latter being $v_\Delta \sim v^2\mu_\Delta /(2M_\Delta^2)$. The masses of the light neutrinos are then given by $m_a = v_\Delta Y^\Delta_{\alpha\beta}U^*_{\alpha a}U^*_{\beta a}$. The smallness of the light neutrino masses is controlled by the smallness of the VEV $v_\Delta$.
Its value is constrained from above by electroweak precision data 
$v_\Delta\leq 4.8\,\text{GeV}$ at 95$\%$ C.L. (see, e.g., \cite{Primulando:2019evb}), and from below by the perturbativity of Yukawa couplings in the generation of light neutrino masses $v_\Delta \gtrsim 0.1\,\text{eV}$ \cite{Cai:2017mow}. 

The phenomenology of the type-II seesaw model is rich due to the presence of seven massive degrees of freedom: three neutral scalars (including the standard Higgs), two singly-charged scalars and two doubly-charged scalars. If the mass of the triplet is at the $\sim \text{TeV}$ scale or below, the new scalars could leave a plethora of signatures at colliders and other low-energy experiments \cite{Atre:2009rg}. Recent studies on type-II seesaw phenomenology and constraints can be found in, e.g., \cite{Primulando:2019evb, Ashanujjaman:2021txz} (see references therein for previous works). 

\subsubsection{Type-III seesaw}
In the type-III seesaw scenario, the SM is augmented with $n_T\geq 2$ hyperchargeless $\text{SU}(2)_L$-triplet fermion fields $\Sigma_{\kappa R}(x)$
\begin{equation}
    \Sigma_{\kappa R}(x) = \begin{pmatrix}
        \Sigma_{\kappa R}^{(0)}(x)/\sqrt{2} & \Sigma_{\kappa R}^+(x)\\
        \Sigma_{\kappa R}^-(x) & -\Sigma^{(0)}_{\kappa R}(x)/\sqrt{2},
    \end{pmatrix}
\end{equation}
$\kappa = 1,\,...,\,n_T$. These fields possess Majorana mass terms and couple to the Higgs and charged lepton doublets through a Yukawa interaction:
\begin{equation}
-\mathcal{L}_\text{Type-III}(x) = \left(Y^\Sigma_{\alpha \kappa} \overline{\psi_{\alpha L}}(x) \Sigma_{\kappa R}(x)i\sigma_2 \Phi^*(x) + \text{h.~c.}\right) + \frac{1}{2}M_{\Sigma_\kappa}\text{Tr}\left[\overline{\Sigma^c_k}(x)\Sigma_\kappa(x)\right],
\end{equation}
with $\Sigma_\kappa(x) = \Sigma_{\kappa R} + \Sigma^c_{\kappa L}(x)$ and $\Sigma^c_{\kappa L}(x) = i\sigma_2 C(\overline{\Sigma_R}(x))^Ti\sigma_2$. The fermion triplets have also gauge interactions, which we do not report here for brevity. 

After the breaking of the electroweak symmetry, the generation of the non-zero neutrino masses proceeds exactly as in the type-I seesaw mechanism via the Yukawa term, with the neutral components $\Sigma_{\kappa R}^{(0)}$ playing the roles of RH neutrinos and a seesaw relation for the light neutrino masses. 

The phenomenology of the type-III seesaw is similar to that in the type-I framework, but richer due to the triplet nature. The neutral components of the triplets mix with the light neutrinos, akin the RH neutrinos in the type-I seesaw scenario, through which they gain couplings to the SM particles. In addition, the fermionic triplets couple to the massive gauge bosons directly and their charged components mix with the other charged leptons. If the masses of the triplets are sufficiently small, there is the possibility to produce them copiously at colliders and observe their indirect signatures \cite{Atre:2009rg} (see, e.g., \cite{Das:2020uer, Ashanujjaman:2021jhi} for recent discussions).

\subsubsection{Leptogenesis within the seesaw extensions of the Standard Model}
A remarkable feature of the seesaw extensions of the SM is the possibility to explain the present \textit{baryon asymmetry of the Universe} (BAU) through the mechanism of \textit{leptogenesis} (LG) \cite{Fukugita:1986hr}. In presence of $L$-, C- and CP-violating processes (e.g., decays or scatterings) allowed by the seesaw models, when these occur out-of-equilibrium in the expanding Universe, all the three Sakharov's condition for a dynamical generation of a lepton asymmetry are fulfilled \cite{Sakharov:1967dj}. Once a lepton asymmetry is generated in the early Universe, it gets converted into the present BAU by the non-perturbative sphaleron processes \cite{Kuzmin:1985mm} 
 predicted by the SM 
 to be in thermal equilibrium above the electroweak scale
\cite{DOnofrio:2014rug}. 

\begin{figure}
\centering
    \includegraphics[width = .9\textwidth]{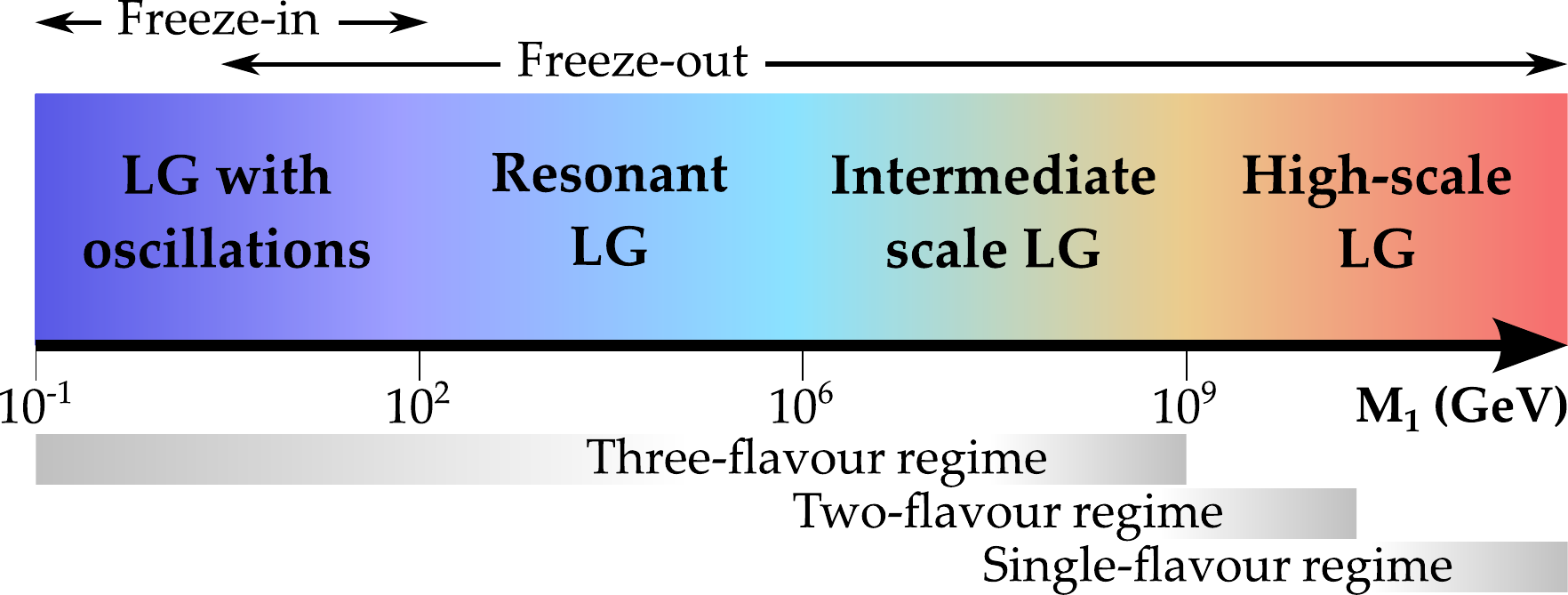}
    \caption{Scheme of LG scenarios within the type-I seesaw model. LG operates through either a \textit{freeze-in} or \textit{freeze-out} mechanism, if the BAU is generated when heavy Majorana neutrinos are being produced or depart from thermal equilibrium, respectively. The single-, two-, and three-flavour regimes are based on how many lepton flavours are distinguishable in the primordial plasma \cite{Nardi:2006fx,Abada:2006fw}. High-scale LG \cite{Fukugita:1986hr, Buchmuller:2004nz} is typically explored in the strong hierarchical limit $M_1\ll M_2\ll M_3$ and suffers from the Davidson-Ibarra bound $M_1\gtrsim 10^9$ GeV \cite{Davidson:2002qv}, reducible to $M_1\gtrsim 10^6$ GeV when milder hierarchies and flavour effects are considered \cite{Racker:2012vw, Moffat:2018wke}. Resonant enhancement of the CP-asymmetry in the decay of two nearly degenerate heavy Majorana neutrinos can reduce the LG scale down to the electroweak range \cite{Pilaftsis:2003gt}. In freeze-in LG taking place at sub-TeV scales, CP-violating RH neutrino oscillations can contribute substantially to the BAU generation \cite{Akhmedov:1998qx}.}\label{fig:LG_scheme}
\end{figure}

The scenario of LG within the minimal type-I seesaw model is by far its most well-studied realisation. It can proceed thanks to decays, inverse decays, scatterings, oscillations involving the RH neutrinos, the Higgs and the LH leptons (and sphalerons) when these happen out-of-equilibrium in the expanding Universe. LG within the type-I seesaw model has been shown to work over wide ranges of masses, with different contributions that dominate the BAU generation depending on the scale and heavy neutrino mass spectrum (see, e.g., \cite{Bodeker:2020ghk} for a recent review on the topic). We depict in Fig.~\ref{fig:LG_scheme} a schematic summary of the various LG versions, see the caption for details and references. 

The low-energy LG realisations with quasi-degenerate-in-mass heavy Majorana neutrinos at the sub-TeV scales are compatible with heavy neutrino masses and mixing in the sensitivity ranges of planned and proposed experiments looking for heavy neutrino signatures \cite{Klaric:2021cpi}  (see also \cite{Abada:2018oly}). A relatively large part of the parameter space associated to successful LG will be scrutinised in the future by collider searches, beam-dump and kaon decay experiments, possibly at future discussed colliders \cite{Klaric:2020phc, Drewes:2021nqr}, and by experiments looking at charged lepton flavour violating processes \cite{Granelli:2022eru, Urquia-Calderon:2022ufc}. However, it remains challenging to test the high-scale LG scenarios in which the heavy Majorana neutrinos have masses above the TeV scale. Nonetheless, an intriguing possibility arises when the requisite CP-violation in LG is uniquely provided by the Dirac and/or Majorana phases of the PMNS matrix. If this is the case, there is a direct link between the BAU and CP-violating phenomena in low-energy neutrino oscillations and $(\beta\beta)_{0\nu}$-decay. The possibility of having CP-violation only from the PMNS matrix can be achieved if the CI matrix is CP-conserving, as, e.g., in flavour models based on sequential dominance \cite{King:2006hn} or with residual CP-symmetries \cite{Chen:2016ptr, Hagedorn:2016lva}. LG with CP-violation solely from the PMNS phases has been shown to work both in high-scale and low-scale scenarios \cite{Pascoli:2006ci, Anisimov:2007mw, Molinaro:2009lud, Molinaro:2008cw, Dolan:2018qpy, Moffat:2018smo, Granelli:2021fyc, Granelli:2023tcj}. If, in the future, LBL neutrino experiments and, possibly, experiments on $(\beta\beta)_{0\nu}$-decay, determine with confidence that the Dirac and/or the Majorana phases of the PMNS matrix have CP-violating values, that would be an indication in favour of LG as the origin of the predominance of matter over antimatter in our Universe. 

Finally, we briefly comment on the LG realisation within the type-II and type-III seesaw models. Unlike the type-I seesaw scenario, LG within the minimal realisation of the type-II seesaw mechanism is unable to reproduce the present BAU \cite{Hambye:2012fh}. To render LG successful, it takes either additional scalar triplets \cite{Ma:1998dx}, support from RH neutrinos \cite{Hambye:2003ka}, or incorporation of the Affleck-Dine mechanism \cite{Affleck:1984fy} with the neutral component $\Delta^{(0)}$ playing the role of the inflaton \cite{Barrie:2021mwi, Barrie:2022cub}.
LG within the type-III seesaw works qualitatively as in the type-I scenario, with more complications caused by the gauge interactions of the charged fermionic triplets, which tend to bring them into thermal equilibrium faster \cite{Hambye:2003rt, Strumia:2008cf}.

\section{The origin of the lepton mixing pattern}\label{sec:LepmixOrigin}
 Understanding the origin of the neutrino mixing pattern and of the light neutrino masses, along with the origin and magnitude of the CP-violation in the leptonic sector, represents a fundamental task in neutrino physics.
 The structure of the PMNS matrix, derived from the neutrino oscillation data, and the spectrum of the lepton masses may be completely random \cite{deGouvea:2003xe, deGouvea:2012ac, Hall:1999sn} or hiding an underlying (approximate) symmetry. The latter idea is based, essentially, on the following logic. Fits to oscillation data give values of $\theta_{12}$ and $\theta_{23}$ which are relatively large compared to $\theta_{13}$, with $\theta_{23}$ being quasi-maximal (i.e., $\sin 2\theta_{23}\sim 1$), $\theta_{13}$ somewhat close to zero and $\theta_{12}$ not distant from special values (e.g., $\sin^2 \theta_{12}\sim 1/2, 1/3$). This structure closely resembles specific patterns like the tri-bimaximal with $\sin^2 \theta_{12} = 1/3$, $\sin 2\theta_{23} = 1$ and $\sin \theta_{13} =0$ \cite{Harrison:2002er, Xing:2002sw}, the bi-maximal with $\sin 2 \theta_{12} = \sin 2\theta_{23} = 1$ and $\sin \theta_{13} =0$ \cite{Vissani:1997pa, Barger:1998ta}, or other alternatives involving the golden ratio \cite{Datta:2003qg, Everett:2008et, Kajiyama:2007gx, Rodejohann:2008ir, Adulpravitchai:2009bg} or the hexagonal mixing \cite{Albright:2010ap, Kim:2010zub}. Deviations from such specific patterns can arise after the breaking of certain symmetries in the flavour sector, like those based on the action of discrete non-Abelian groups such as $S_N$, $A_N$, $D_N$, $\Sigma(2N^2)$, $\Sigma(3N^3)$, $\Delta(3N^2)$, $\Delta(6N^2)$ (see, e.g., \cite{Ishimori:2010au} for the definitions and properties of these groups and, e.g., \cite{Altarelli:2010gt, King:2013eh, Petcov:2017ggy, Feruglio:2019ybq} for reviews). Models based on such discrete flavour symmetries to describe the lepton mixing pattern are highly predictive and establish correlations between the values of the PMNS mixing angles and the CP-violating phases. It is also possible to combine the discrete flavour symmetry approach with a generalised CP symmetry \cite{Feruglio:2012cw, Holthausen:2012dk} to further reduce the number of free parameters and increase the correlations with the PMNS phases. Future high-precision determination of the mixing parameters will be crucial for discriminating between the models considered in the literature \cite{Agarwalla:2017wct, Blennow:2020snb, Blennow:2020ncm}.

 A possible drawback of such models is that the correct breaking of the discrete flavour symmetry generically implies the existence of many scalar fields, called \textit{flavons} \cite{Froggatt:1978nt}, entering an elaborate scalar potential with complicated symmetric shaping. Recently, the idea of \textit{modular invariance} has been proposed to provide substantial simplification to this problem \cite{Feruglio:2017spp}. In this approach, the flavour (modular) symmetry is broken by the non-zero vacuum expectation value of a single scalar field, the \textit{modulus}, the vacuum expectation value of which can also be the only source of violation of the CP-symmetry. The Yukawa couplings and the fermion mass terms in the theory are represented by special functions of the modulus called \textit{modular forms} having particular properties under the modular group transformations. Over less than a decade, the number of works on the modular invariance solution to the flavour problem has grown enormously (see, for instance, \cite{deMedeirosVarzielas:2023crv} and extensive lists of references therein) and the current hope is to explain under this same formalism the masses and mixing patterns of both leptonic and quark sectors.

\section{Outlook and Conclusion}\label{sec:Summary}
The parameters of the electroweak theory and most of the ones of the three-neutrino mixing paradigm are known today with exceptional precision. Nevertheless, the following fundamental tasks persist in neutrino physics.

\begin{itemize}
\item What are the ordering and the absolute mass scale of neutrino masses? 
\item What is the amount of CP-violation in the lepton sector? 
\item Are neutrinos Dirac or Majorana particles? 
\item What are the precise values of the mixing angles? \item Is the three-neutrino mixing accurate or are there deviations, such as sterile neutrinos, non-unitary PMNS matrix and/or non-standard interactions? 
\item What mechanism is responsible for the generation of light neutrino masses?
\item What is the origin of the lepton masses and mixing pattern, and why is it substantially different from that in the quark sector?
\end{itemize}

The above questions form the basis of an ambitious research programme and motivate extensions of the Standard Model that address the origin of neutrino masses and mixing. Ongoing and upcoming neutrino experiments on neutrino oscillations aim to refine the neutrino mixing picture, determining with more precision the parameters that have already been measured, unveiling those that are currently unknown, revealing potential correlations among them, and identifying/constraining deviations from the three-neutrino mixing paradigm. Additionally, planned and proposed accelerator-based and low-energy experiments will explore more in depth the parameter spaces of low-energy realisations of seesaw models, which constitute attractive mechanisms of light neutrino mass generations.

Resolving these central issues in neutrino physics may have profound implications also for other fundamental problems in modern physics, such as the matter-antimatter asymmetry of the Universe and the nature of dark matter. Moreover, the nascent field of multi-messenger astronomy involving neutrinos offers further avenues for exploration. Hopefully, in the foreseeable future, some, if not all, of these lingering questions in neutrino physics will be clarified.

\section*{Acknowledgements}
We thank Silvia Pascoli and Serguey T. Petcov for useful comments, as well as Jin-Wei Wang and Jaime Hoefken Zink for proof-reading this manuscript. This work was supported in part by the European Union's Horizon research and innovation programme under the Marie Skłodowska-Curie grant agreements No.~860881-HIDDeN and No.~101086085-ASYMMETRY, and by the Italian INFN program on Theoretical Astroparticle Physics.


\providecommand{\href}[2]{#2}\begingroup\raggedright\endgroup

\end{document}